# Automated Coding of Communications in Collaborative Problem-solving Tasks Using ChatGPT

Jiangang Hao[*], Wenju Cui, Patrick Kyllonen, Emily Kerzabi, Lei Liu, and Michael Flor

ETS Research Institute, Princeton, NJ 08541, USA

https://arxiv.org/abs/2411.10246

**Abstract**

Collaborative problem solving (CPS) is widely recognized as a critical 21st-century skill. Assessing CPS depends heavily on coding the communication data using a construct-relevant framework, and this process has long been a major bottleneck to scaling up such assessments. Based on five datasets and two coding frameworks, we demonstrate that ChatGPT can code communication data to a satisfactory level, though performance varies across ChatGPT models, and depends on the coding framework and task characteristics. Interestingly, newer reasoning-focused models such as GPT-o1-mini and GPT-o3-mini do not necessarily yield better coding results. Additionally, we show that refining prompts based on feedback from miscoded cases can improve coding accuracy in some instances, though the effectiveness of this approach is not consistent across all tasks. These findings offer practical guidance for researchers and practitioners in developing scalable, efficient methods to analyze communication data in support of 21st-century skill assessment.

---

[*] Email: jhao@ets.org

# Introduction

Collaborative Problem Solving (CPS) has been widely recognized as a critical 21st-century skill essential for success in education and the workforce (OECD, 2017). CPS encompasses the ability to work effectively with others to solve complex problems by exchanging ideas, negotiating solutions, and constructing shared understandings. It is a core competency in STEM fields, business, healthcare, and many other domains where individuals must collaborate to generate innovative solutions. Assessing CPS has gained growing attention in educational measurement, given its central role in fostering the development of CPS skills.

Verbal communication data is crucial for assessing CPS, as it contains rich information about the group interactions, learning dynamics, and knowledge construction processes. Verbal communication can take various forms, such as text chats and audios. With the rapid advancement of AI technology, audio can be automatically transcribed into text with high accuracy (Radford et al., 2023). Therefore, the primary focus of analysis is on the communication data in text format. A common approach to extracting evidence from text is to code (classify) communication text, such as conversational turns, into categories based on predefined frameworks or rubrics that align with the constructs of interest. Historically, this coding process has leaned heavily on the painstaking efforts of human raters, who must meticulously categorize each turn or multiple turns of chats into some predefined categories based on the coding framework. Once a portion of the data has been reliably human-rated (e.g., 10 - 25% of the total dataset; Campbell et al., 2013), automated coding methods can be developed using numerical representations of the chats (e.g., n-grams or neural embeddings) and supervised machine learning classifiers (Flor et al., 2016; Hao et al., 2017a; Moldovan et al., 2011; Rosé et al., 2008). Nevertheless, this pipeline hinges on high-quality human-coded data, which remains a time-consuming and labor-intensive part of the process despite constituting only a fraction of the total dataset. This inherent limitation significantly constrains the scope and scalability of research and operations in this area.

As the volume of digital communication data continues to grow, there is a pressing need for more efficient methods to automatically code large datasets without compromising the depth and quality. Recent advances in large language models (LLMs) and generative AI have opened new possibilities for automated scoring and coding (Hao et al., 2024). For example, ChatGPT is an AI-powered conversational agent developed by OpenAI (2023), designed to engage in human-like dialogue. It uses LLMs, such as GPT-4 and GPT-4o, to understand and generate natural language responses, enabling it to assist with various tasks like answering questions, generating text, and facilitating interactive conversations. It is, therefore, tantalizing to consider the possibility of leveraging ChatGPT as a substitute for human raters in coding communication data.

Traditional automated coding methods rely heavily on large amounts of human-coded data to train supervised machine learning models. In contrast, ChatGPT can be directly instructed to apply various coding frameworks to code chat data without extensive human involvement. This approach aligns with zero-shot or few-shot classification in natural language processing (NLP), where the model can perform tasks with little to no prior labeled data. While the concept is straightforward, the true measure of its effectiveness lies in its performance in real-world scenarios, which is tied to the complexity of the data and the coding framework employed. Studies have indicated that ChatGPT generally exhibits good accuracy for straightforward tasks like sentiment analysis, albeit with some variance across datasets (Belal et al., 2023; Fatouros et al., 2023). However, when confronted with more complex coding tasks, ChatGPT often falls short of meeting expectations (Casabianca et al., 2025; Kocoń et al., 2023). It is, therefore, an empirical question whether we can instruct ChatGPT to code the communication from collaborative problem-solving tasks accurately. This paper presents an empirical study that seeks to address four research questions that are critical for understanding the applicability of ChatGPT-based automated coding in assessing CPS:

**RQ1.** How accurately can different ChatGPT models code chat data from CPS tasks?

**RQ2.** How does the nature of communication from different CPS tasks influence ChatGPT's coding performance?

**RQ3.** How does the choice of coding framework affect the accuracy of ChatGPT coding?

**RQ4.** Can we improve ChatGPT coding performance by incorporating feedback from miscoded cases into the prompt?

Based on data from five CPS tasks and two coding frameworks, we provide empirical evidence that using ChatGPT to code text-based chat data is feasible and effective. This approach significantly reduces the time and cost associated with manual coding, offering a promising pathway for accelerating and scaling up research that relies on analyzing extensive communication data.

## Literature Review

Collaboration is defined as a "coordinated, synchronous activity that results from a continued attempt to construct and maintain a shared conception of a problem." (Roschelle & Teasley, 1995). CPS is a special form of collaboration that integrates both cognitive and social processes, involving two or more individuals who interact to share and negotiate ideas, coordinate learning activities, and apply social strategies to sustain engagement while solving a shared problem (Dillenbourg et al., 2009; Järvelä et al., 2010; Liu, et al., 2015; Van den Bossche et al., 2006). In PISA 2015 (OECD, 2017, p. 134, Box 7.1) collaborative problem solving competency was formally defined as "the capacity of an individual to effectively engage in a process whereby two or more agents attempt to solve a problem by sharing the understanding and effort required to come to a solution and pooling their knowledge, skills, and efforts". CPS offers several advantages over individual work, including a more effective division of labor, broader integration of knowledge and perspectives, and enhanced creativity through idea exchange (OECD, 2017). As such, CPS competency is widely recognized as a critical 21st-century skill essential for success in education and the workforce (Fiore et al., 2017; Griffin et al., 2012; OECD, 2017; Roschelle & Teasley, 1995; World Economic Forum, 2025).

Research on CPS has mainly focused on its role in learning, such as designing effective collaborative learning environments (Koschmann, 1996; Stahl et al., 2006) or developing tasks to foster collaboration skills to improve learning outcomes (Sottilare et al., 2012). In contrast, the assessment of CPS has been relatively underexplored, primarily due to its intrinsic complexity, high logistical costs, and the challenges of ensuring measurement qualities such as validity, reliability, and fairness (Hao et al., 2017b). The constructs underlying CPS are complex and multidimensional, requiring assessment designs that balance the coverage of the constructs with practical feasibility. On one hand, there is a strong desire to capture CPS in its full breadth, incorporating cognitive, social, and behavioral dimensions. On the other hand, practical constraints, including time, budget, and psychometric rigor, necessitate tradeoffs in assessment design (von Davier et al., 2017).

The advancement of digital technology has enabled computer-mediated online collaboration, making it feasible to conduct large-scale collaboration, which is a prerequisite for assessing CPS at scale. As a result, two notable survey assessments of CPS have been developed, the Assessment and Teaching of 21st-Century Skills (ATC21S) (Griffin et al., 2012) and the 2015 Programme for International Student Assessment (PISA 2015) (OECD, 2013; Graesser et al., 2018). ATC21S assessed CPS by pairing two students who collaborate via text chat to complete tasks. The scoring was based on students' task responses and actions (Hesse et al., 2015; Scoular et al., 2017). Despite the chat communication containing rich data on CPS, the communication data were not explicitly scored due to technology and cost constraints, limiting the assessment's ability to tap into the key social aspects of CPS. Additionally, the scoring model did not consider the interdependence between partners' problem-solving proficiency and CPS skills (Hao et al., 2019).

PISA 2015 adopted a different approach by placing students in teams with virtual partners programmed to simulate collaboration in a standardized manner (Graesser et al., 2017; He et al., 2017). Students interacted with these virtual agents through predefined text responses designed by experts, ensuring consistency across test administrations. While this approach enhances standardization and reliability, it does not fully reflect the complexities of real-world collaboration, where communication is dynamic and significantly shaped by the unfolding context of prior exchanges. Both ATC21S and PISA 2015 were significantly constrained by technological limitations, psychometric requirements, and the substantial resources needed for implementation. The cost of developing assessment tasks and scoring turn-by-turn communication makes it prohibitively expensive for testing programs to administer these assessments routinely in a standardized format (Hao et al., 2017b).

On the other hand, testing companies, such as Educational Testing Service (ETS), have conducted systematic research to explore the feasibility of large-scale, standardized Collaborative Problem Solving (CPS) assessments. Research activities centered on defining operationalizable CPS constructs (Andrews et al., 2017; Andrews-Todd & Kerr, 2019; Liu et al., 2015), developing technological infrastructure (Hao et al., 2017c), building assessment prototypes (Andrews-Todd et al., 2019; Hao et al., 2015; Martin-Raugh et al., 2020), and advancing psychometric methodologies (Halpin et al., 2017; Hao et al., 2016; Hao & Mislevy, 2019; Zhu & Andrews-Todd, 2019; Zhu & Zhang, 2017) and automated scoring systems (Flor et al, 2016; Hao, et al, 2017). These research efforts reached a consensus that a key step in assessing CPS is coding the process data (both actions and communications) generated during collaboration. This involves applying a coding framework that identifies key dimensions of CPS skills to categorize participants' actions and communications. These categorized or coded data can then be used to develop higher-level scores of CPS through psychometric modeling.

Over the past decade, multiple coding frameworks have been developed to capture different aspects of CPS for assessment purposes. The CPS framework from ATC21S defined five core skills essential to effective collaboration: participation, perspective-taking, social regulation, task regulation, and knowledge building (Hesse et al., 2015). The PISA 2025 framework adopted a slightly different approach, focusing on three primary CPS competencies: establishing and maintaining shared understanding, taking appropriate actions to solve problems, and organizing team processes effectively (OECD, 2013). These frameworks emphasized CPS's social and cognitive dimensions, though with varying emphases on interaction dynamics and problem-solving strategies.

Other frameworks have taken different perspectives on CPS assessment Liu et al. (2015) proposed a framework centered on the problem-solving process, identifying four key aspects: sharing ideas, negotiating ideas, regulating problem solving, and maintaining communication. This approach provides a structured way to evaluate how teams navigate problem-solving tasks collaboratively. Andrews-Todd et al. (2017) introduced a framework that categorizes team interaction patterns, distinguishing between collaborative, cooperative, dominant-dominant, dominant-passive, expert-novice, and even instances of fake collaboration. This perspective highlights the varying dynamics within groups and their impact on CPS effectiveness. Andrews-Todd and Kerr (2019) developed a framework that distinguishes between social and cognitive dimensions of collaboration. The social dimension includes maintaining communication, sharing information, establishing shared understanding, and negotiating, while the cognitive dimension encompasses exploring and understanding, representing and formulating, planning, executing, and monitoring. More recently, Kyllonen et al. (2023) developed a framework intended to be applied to different types of collaborative tasks. This framework identifies CPS behaviors such as maintaining communication, staying on track, eliciting and sharing information, and acknowledging partners' responses. It emphasizes generalizable interaction patterns and provides a framework to assess CPS across diverse task contexts.

Furthermore, the National Assessment of Educational Progress (NAEP) 2026 mathematics framework integrates CPS to better reflect the collaborative nature of real-world problem solving in mathematics. The framework emphasizes key collaborative skills in mathematics, such as attending to and interpreting others' mathematical contributions, evaluating the validity of peers' ideas, and responding constructively to others' reasoning (National Assessment Governing Board, 2023). The diversity of these frameworks reflects the complexity of collaborative problem solving (CPS) and the wide range of skills required for effective collaboration. While some frameworks emphasize problem-solving processes or interaction patterns, others focus on communication strategies or domain-specific applications, such as mathematics. Despite these differences, they share common foundations in identifying key competencies essential for CPS assessment. Together, they contribute to a more comprehensive competency model for evaluating CPS skills.

Applying these coding frameworks to collaborative task data presents practical and logistical challenges. Typical collaborative tasks generate both process data (e.g., such as participant actions and communications) and outcome data (Hao et al., 2019). These frameworks are primarily used to categorize the process data. While actions are generally drawn from a finite set and thus relatively straightforward to code, communication data is far more complex, rich, varied, and generated in large volumes, making it significantly more challenging to code. Prior to the development of computer-based natural language processing (NLP) technologies, human raters were trained to code these data. This process is highly labor-intensive and expensive to scale.

Advances in natural language processing (NLP) have enabled the development of automated scoring and coding systems (see Flor & Hao, 2021, for a concise review). The typical workflow usually begins by constructing a training dataset in which human raters code a subset of the communication data (often 10% to 25% of the total dataset, Campbell et al., 2013) based on a predefined coding framework. Each communication turn (e.g., a chat message) in the training data is then transformed into a numerical representation using NLP techniques, such as bag-of-words, n-grams (Jurafsky & Martin, 2019), or more recently, neural embedding vectors (e.g., Devlin et al., 2018; Mikolov et al., 2013; Pennington et al., 2014). Finally, statistical models or supervised machine learning classifiers are trained to learn the relationship between these numerical representations and the coding categories assigned by human raters (Flor et al., 2016; Flor & Andrews-Todd, 2022; Hao et al., 2017a; Moldovan et al., 2011; Rosé et al., 2008). In recent years, the emergence of LLMs has enabled end-to-end approaches for mapping communication data directly to coding categories through model fine-tuning. Studies have reported that fine-tuned LLMs often outperform traditional machine learning models' coding accuracy (Zhu et al., 2024). While such automated coding approaches significantly reduce the need for extensive human coding, the initial development of a training dataset remains labor-intensive. This requirement poses a substantial barrier to exploring multiple coding frameworks on the same dataset.

Since the release of ChatGPT in late 2022, chat LLMs have shown remarkable progress in aligning LLM responses with human queries. This raises the possibility that we may skip the human coding process altogether by directly instructing LLMs with the coding framework to code data. Studies have indicated that ChatGPT generally exhibits good accuracy for straightforward tasks like sentiment analysis, albeit with some variance across datasets (Belal et al., 2023; Fatouros et al., 2023). However, when confronted with more complex coding tasks, ChatGPT often falls short of meeting expectations (Kocoń et al., 2023). This raises an important empirical question: Can ChatGPT be effectively prompted to code communication data from collaborative problem-solving tasks? To address this question, we carried out a systematic empirical investigation, the methodology and findings of which are presented in the following sections.

## Methods

In this section, we introduce the methods employed in this study. Specifically, we outline the study design to address the research questions, detailing the collaborative tasks, data, coding framework, LLMs, and prompt design, as well as how to improve the prompt with feedback.

**Study Design**

This study was designed to address the key research questions outlined in the introduction. To investigate Research Question 1 (*how accurately can different ChatGPT models code chat data from CPS tasks?*), we selected four leading LLMs underlying ChatGPT: GPT-4, GPT-4o, GPT-o1-mini, and GPT-o3-mini. Each model was prompted using the same instructions and applied to the same dataset, allowing for a direct comparison of their coding performance.

To address Research Question 2 (*how does the nature of communication from different CPS tasks influence ChatGPT's coding performance?*), we selected communication data from five distinct collaborative tasks and compared the performance of the same LLMs across them. In addition, we examined whether the presence of scientific terminology influences ChatGPT coding accuracy, given that domain-specific terms have been associated with lower-quality embeddings compared to more commonly used vocabulary (Beltagy et al., 2019; OpenAI, 2023).

To address Research Question 3 (*how does the choice of coding framework affect the accuracy of ChatGPT coding?*), we selected two different coding frameworks, one grounded primarily in theory, and the other informed by both theoretical foundation and empirical data. Each framework has been applied to chat data from multiple collaborative tasks. For Research Question 4 (*can we improve ChatGPT coding performance by incorporating feedback from miscoded cases into the prompt?),* we identified frequently miscoded categories, revised the prompts accordingly, and re-applied them to code the data.

The following subsection provides details on the datasets, coding frameworks, LLMs, prompt design, evaluation methods and the strategies for diagnosing and improving low coding performance.

**Datasets**

The dataset used in this study was drawn from five CPS tasks. Two of these tasks involved problem-solving in scientific contexts, one focused on condensation and the other on volcanic eruptions. For brevity, we refer to them hereafter as the *Condensation* and *Volcano* tasks, respectively. In the *Condensation* task, each team comprises two participants who collaborate via text chat to answer a series of questions about how condensation forms and the factors that influence it (Liu et al., 2013). Similarly, in the *Volcano* task, pairs of participants work together through text-based communication to explore how seismometers can be used to predict volcanic eruptions (Hao et al., 2015). Figure 1 presents screenshots of both tasks.

**Figure 1.**

*Screenshots of the Two Science Tasks*.

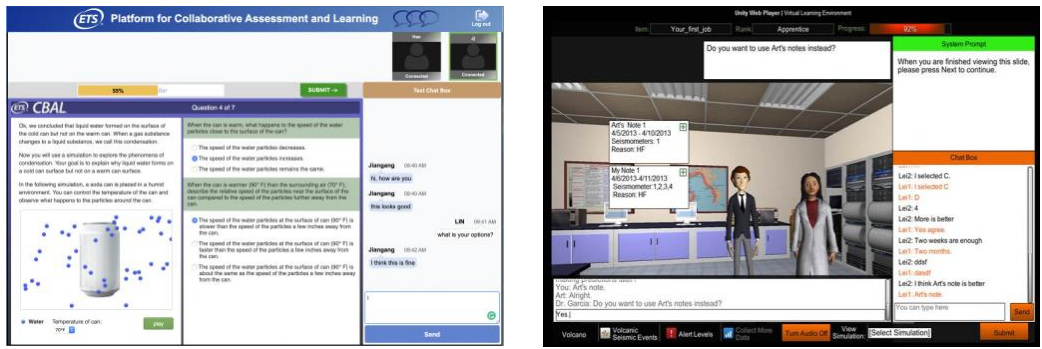

*Note*. Left: *Condensation* task. Right: *Volcano* task.

The remaining three collaborative tasks focus on three general cognitive skills—*Negotiation*, *Decision-Making*, and *Letter-to-Number* (Kyllonen et al., 2023). Each of these tasks involved a team of four participants collaborating online via text chat. In the *Negotiation* task, team members worked together to plan a fundraising event. Each participant received a list of options, with varying payoffs depending on the individual. Because identical options yielded different payoffs for different participants, the team was to negotiate to reach a mutually agreed solution that maximized individual gains without causing the negotiation to break down, an outcome that would result in no payoff for anyone (Martin-Raugh et al., 2020). In the *Decision-Making* task, four team members engaged in a text-based discussion to select the most suitable apartment from a set of candidates. Each apartment had its own advantages and disadvantages, but each participant only saw a subset of the attributes. To make an informed group decision, team members were to share their unique information with one another. In the *Letter-to-Number* task, participants collaborated to uncover a hidden mapping between letters and numbers, which is classic reasoning task (Newell & Simon, 1972). Each letter was associated with a predefined number (e.g., A = 1, B = 3, C = 4), and participants propose operations such as A + B. The system returns the result based on the numerical values of the letters (e.g., A + B = 1 + 3 = 4, so the system responds with C). Figure 2 presents screenshots of these three tasks.

**Figure 2.**

*Screenshots of the three general cognitive skill tasks*.

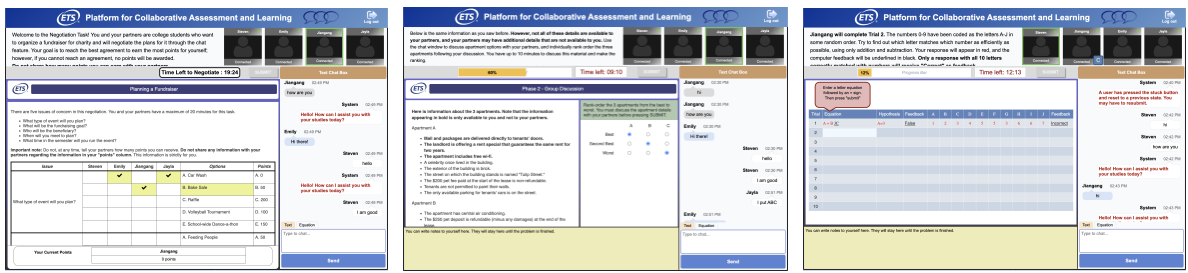

*Note*. Left: *Negotiation* task. Middle: *Decision-Making* task. Right: *Letter-to-Number* task.

Data were collected through crowdsourcing platforms. Specifically, the data for the *Volcano* and *Condensation* tasks were collected from Amazon Mechanical Turk

(https://www.mturk.com/) and the data for the other three tasks were collected through Prolific (https://www.prolific.com/). All data collections were reviewed and approved by the Institutional Review Board (IRB) of Educational Testing Service (ETS). Each collaborative task session lasted approximately 40 minutes and generated an average of 50 to 100 chat turns. For this study, we randomly sampled a subset of completed collaborative sessions from each task, resulting in approximately 1,500 chat turns for each task.

**CPS Coding Frameworks**

We used two coding frameworks in this study. Coding Framework 1 (Liu et al., 2016) was developed based on an extensive review of research in computer-supported collaborative learning (CSCL) (Barron, 2003; Dillenbourg & Traum, 2006; Griffin & Care, 2014), along with the PISA 2015 Collaborative Problem Solving (CPS) Framework (Graesser & Foltz, 2013). It includes four major categories designed to capture key social and cognitive processes within group interactions. We applied this framework to the chat data from the *Volcano* and *Condensation* science tasks. Coding Framework 2 (Kyllonen et al., 2023) was developed through a more data-driven approach by combining elements of Framework 1 with a negotiation-specific framework (Martin-Raugh et al., 2020). This framework was applied to the three tasks focused on general cognitive skills.

Each framework was implemented by two trained human raters who independently coded each chat turn. Inter-rater reliability was monitored using Cohen's Kappa statistics (Cohen, 1960). It is worth noting that for each task, one of the two human raters is an expert rater based on their expertise and training. In our subsequent comparisons of coding agreement between human raters and ChatGPT models, we use the expert raters' codes as the reference, though the agreement between ChatGPT models and the non-expert raters is very similar to that with the expert raters. Table 1 presents the coding categories and definitions for both frameworks, while Tables 2 and 3 provide sample chat excerpts from the *Volcano* and *Decision-Making* tasks, along with their corresponding human-assigned codes.

**Table 1**

*Two CPS Coding Drameworks Used in this Study*

| Coding Framework 1 | Coding Framework 2 |
|---|---|
| **1. Sharing ideas**: chats that capture instances of how individual group members introduce diverse ideas into collaborative discussions. For example, participants may share their responses to assessment items or highlight pertinent resources contributing to problem solving. | **1. Maintaining communication (MC)**: chats that involve greetings, emotional responses (including emojis), technical discussions, and other communications that cannot be classified elsewhere. |
| **2. Negotiating ideas**: chats that aim to document evidence of collaborative knowledge building and construction through negotiation. Examples include agreement/disagreement, requesting clarification, elaborating or rephrasing others' ideas, identifying cognitive gaps, and revising ideas. | **2. Staying on task (OT)**: chats that keep things moving, that involve monitoring time, and steering team effort. |
| **3. Regulating problem solving**: chats that focus on the collaborative regulation aspect of team discourse, including activities like identifying team goals, evaluating teamwork, and checking to understand. | **3. Eliciting information (EI)**: chats that elicit information from another about the task, including strategies, goals, and opinions. |
| **4. Maintaining communication**: chats that capture content-irrelevant social communications that contribute to fostering or sustaining a positive communication atmosphere within the team. | **4. Sharing information (SI)**: chats that share information, strategies, goals, or opinions. |
| | **5. Acknowledging (AK)**: chats involving acknowledging partners' input, stating agreement or accepting a tradeoff, stating disagreement or rejecting a tradeoff, |

building off one's own or a teammate's idea, and proposing a negotiation tradeoff or suggesting a compromise.

*Note*. Coding Framework 1 was used for *Volcano* and *Condensation*; Coding Framework 2 was used for *Letter-to-Number*, *Decision-Making*, and *Negotiation*.

**Table 2**

*Example Chats and Coding from the Volcano Task*

| Team Member | Chat | Skill Category |
|---|---|---|
| Person A | hi | Maintain |
| Person B | hey | Maintain |
| Person B | Wait you're a real person? | Maintain |
| Person A | haha yea, i guess its just a coincidence | Maintain |
| Person A | Welp, i guess its gonna be a good day | Maintain |
| Person B | Easy 5 dollars | Maintain |
| Person A | so i put that they were both moving | Share |
| Person B | It's either C or A | Share |
| Person B | It's A | Share |
| Person A | Sounds good to me, i think molecules are always moving | Negotiate |
| Person B | Yeah | Maintain |
| Person B | Definitly B | Share |
| Person A | Well i think it depends on the temperature of the water | Negotiate |
| Person A | if the water is all one temp then they are moving at the same speed, or am i wrong | Negotiate |
| Person B | I'm not sure. | Regulate |
| Person A | wanna just go with c | Regulate |

*Note*. Coding categories are defined in Table 1. Maintain = Maintaining Communication; Share = Sharing ideas; Negotiate = Negotiating ideas; Regulate = Regulating problem solving.

**Table 3**

*Example Chats and Coding from the Decision-Making Task*

| Team Member | Chat | Skill Category |
|---|---|---|
| Person A | What does everyone say? | Elicit |
| Person B | Avenue A | Share |
| Person C | B: several billboard arounds town, no clean up crew provided, full time staff known to be surly | Share |
| Person A | Yuck! | Maintain |
| Person A | I have ACB so far? | Elicit |
| Person C | c: option to include live local music bands at a low cost, equipped to host weddings and birthdays, owners like 10 mins from site | Share |
| Person A | For A, it says that there can be multiple events going on at the same time in close proximity to one another | Share |
| Person C | ohh thats a turn off | Share |
| Person A | For C, patrons have to provide their own beverages, but that sounds helpful (cheaper for BYOB) | Share |
| Person B | I think safety and security should be a priority here | Share |
| Person A | B has poor lighting in the parking lot and along outdoor walkways AND customers must contribute to the cost of liability insurance | Share |
| Person A | 2 minutes left | On Track |
| Person A | ACB? or ??? | Elicit |
| Person A | B sounds good for food | Share |
| Person B | ACB would be best option | Share |
| Person A | That's what I had | Acknowledge |

*Note.* Coding categories are defined in Table 1. Elicit = Eliciting Information; Share = Sharing Information; On Track = Staying on Task; Acknowledge = Acknowledging.

**LLMs and Prompt Design**

We selected OpenAI's GPT models deployed on the Azure cloud platform for this study, as they are widely recognized among the most capable large language models (LLMs) available (LMArena, n.d.; LLM Stats, n.d.). Specifically, we used four model variants: GPT-4 (version: *turbo-2024-04-09*), GPT-4o (version: *2024-05-13*), GPT-o1-mini (version: *2024-09-12*), and GPT-o3-mini (version: *2025-01-31*). To ensure consistency and reproducibility, all models were run with the temperature parameter set to zero and a fixed random seed. The *o1-mini* and *o3-mini* models are designed for enhanced reasoning capabilities, often referred to as "thinking" or "reasoning" models, which reportedly produce more coherent and logically consistent responses by internally simulating a chain-of-thought process.

The art of crafting effective prompts for large language models (LLMs) lies at the intersection of creativity and technical skill (Brown et al., 2020). Prior research has shown that incorporating relevant domain knowledge is essential for designing effective prompts (Zamfirescu-Pereira et al., 2023), and that iterative refinement is key to fully leveraging the capabilities of LLMs. As a result, *prompt engineering* has emerged as a specialized discipline, in which practitioners design carefully structured prompts to elicit optimal responses from AI systems (Radford et al., 2023. In this study, we conducted a comprehensive prompt engineering process guided by best practices (OpenAI, n.d.) to develop prompts that instructed the LLMs to code chat messages accurately without being overly prescriptive. Each prompt consisted of four main components: (1) a statement of the task goal, (2) a description of the coding framework, (3) approximately 10 expert-generated examples per category, and (4) specifications for the input and output format, followed by the actual chat data. Appendix A provides sample prompts corresponding to each of the two coding frameworks used in this study. When applying the same framework across different tasks (which we did for the *Letter-to-Number*, *Negotiation*, and *Decision-Making* tasks; and separately for volcano and condensation task), we reused the prompt structure, modifying only the task-specific chat examples.

**Diagnosis of Coding Performance**

We used Cohen's Kappa to evaluate the coding performance of GPT models. Cohen's Kappa is a statistical measure of inter-rater agreement for categorical items, accounting for agreement occurring by chance. It provides a more robust and reliable metric than simple percentage agreement and is widely recognized as a standard for assessing inter-rater reliability in classification tasks (Cohen, 1960). In our study, we used the Kappa coefficient between a human rater and the LLMs to assess the performance of ChatGPT coding.

We observed that agreement between the ChatGPT models and human raters was consistently lower for the two science tasks coded using Framework 1 compared to the 3 tasks coded using Framework 2. To investigate potential causes and identify avenues for improvement, we conducted two follow-up analyses. First, we hypothesized that the presence of scientific terminology may contribute to the reduced performance, due to limited training data and domain-specific terminology use. To test this, we compiled a list of domain-specific scientific terms (see Appendix B) and labeled each chat turn with a 1 if it contained any of these terms, and 0 otherwise. We then calculated Cohen's Kappa separately for the two turn categories and compared the levels of agreement with human raters for each category. Second, we examined whether incorporating feedback on frequently miscoded cases could enhance the ChatGPT models' coding performance. Using cases with full agreement between human raters as ground truth, we identified the most common miscoding made by the best performing GPT-4o model and incorporated feedback about these cases into a revised prompt. We then re-coded the chat data using the updated prompt. If this approach led to improved performance, we extended the analysis to cases where the human raters initially disagreed. These follow-up analyses directly address Research Questions 2 and 4.

## Results

This section presents the results organized by research question. For Research Question 1, the coding performance of different ChatGPT models is summarized in Table 4 and Figure 3. For the three general cognitive skill tasks, *Negotiation*, *Letter-to-Number*, and *Decision-Making*, which were coded using the Coding Framework 2, all ChatGPT models achieved agreement levels with human raters that were comparable to inter-rater agreement between humans. In contrast, for the two science tasks, *Condensation* and *Volcano*, coded using the Coding Framework 1, the agreement between ChatGPT models and human raters was lower than that observed between human raters. This suggests that human raters may be able to detect nuances in the communication based on Coding Framework 1 that the LLMs fail to capture, or that the tasks themselves differ in fundamental ways. As the subsequent analysis will show, this discrepancy is more likely due to the coding framework than to differences in the tasks.

Among all ChatGPT models evaluated, GPT-4o demonstrated the highest overall performance. Notably, the more recent reasoning-oriented models, GPT-o1-mini and GPT-o3-mini, did not outperform GPT-4o. This indicates that the added reasoning capabilities of these newer models may not translate into better performance on this particular coding task. These findings offer practical guidance for selecting GPT models in similar applications to optimize cost and efficiency.

**Table 4**

*Agreement (Cohen's Kappa) on Five Different Collaborative Tasks*

| Task | Human vs. human | Human vs. GPT-4 | Human vs. GPT-4o | Human vs. GPT-o1-mini | Human vs. GPT-o3-mini | # of turns |
|---|---|---|---|---|---|---|
| Condensation | 0.779 | 0.525 | 0.576 | 0.457 | 0.493 | 1509 |
| Volcano | 0.685 | 0.554 | 0.604 | 0.518 | 0.569 | 1510 |
| *Negotiation* | 0.527 | 0.576 | 0.612 | 0.561 | 0.574 | 1513 |
| Letter-to-Number | 0.739 | 0.735 | 0.728 | 0.728 | 0.733 | 1508 |
| *Decision-Making* | 0.683 | 0.733 | 0.694 | 0.687 | 0.729 | 1510 |

*Note.* Turns are chat conversational turns based on randomly sampled collaborative sessions from a larger set of sessions.

**Figure 3**

*The Graphic Presentation of the Coding Agreement Results as Shown in Table 4.*

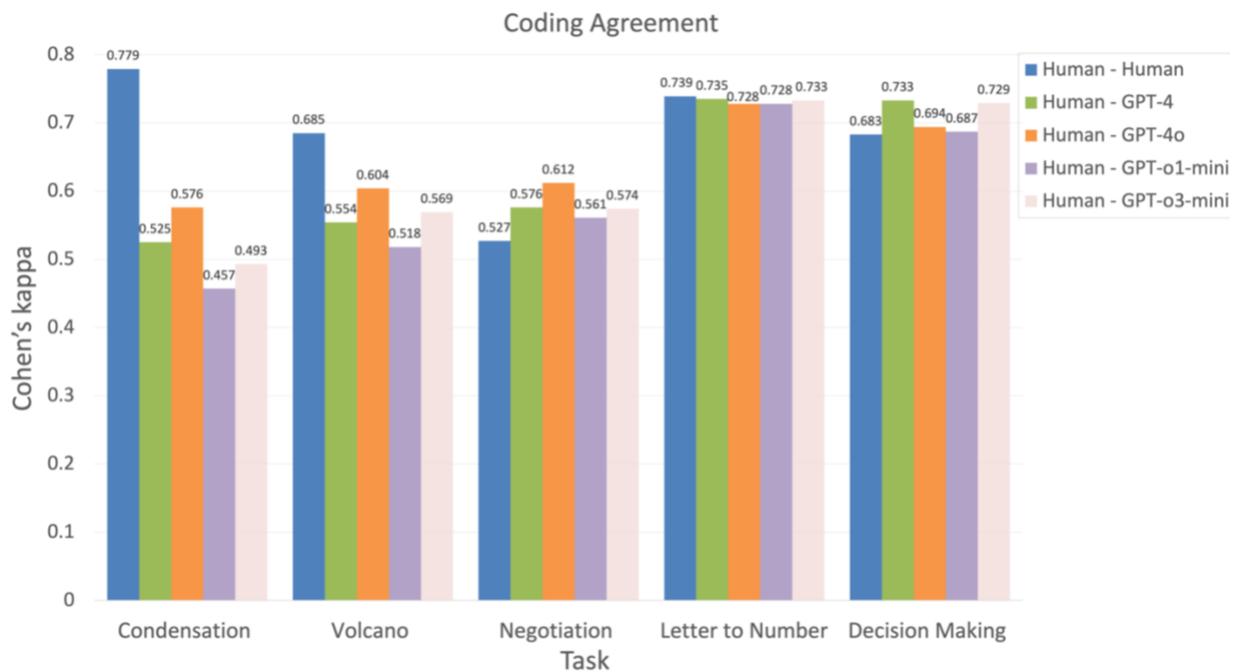

For Research Question 2, we observed that coding agreement varied by task, although similar patterns emerged across ChatGPT models. Specifically, tasks with lower inter-rater agreement between human raters also tend to exhibit lower agreement between human raters and ChatGPT models. Among the three general cognitive skill tasks that share the same coding framework 2, the *Negotiation* task yielded the lowest coding agreement for both human-human and human-ChatGPT comparisons. These findings underscore that the content and nature of communication significantly influence the performance of ChatGPT models on coding tasks.

Another notable pattern was the consistently lower performance of ChatGPT on the two science-related tasks under the coding framework 1. This prompted the hypothesis that the presence of domain-specific scientific terminology may contribute to the reduced coding accuracy. To test this hypothesis, we separated chat turns from the two science tasks into two groups: those containing scientific terms (as listed in Appendix B) and those that did not. We then computed coding agreement for each group using the best-performing model, GPT-4o. Table 5 presents the results.

Two key observations emerged from Table 5. First, chat turns containing scientific terminology showed lower coding agreement for both human raters and GPT-4o, which might suggest that such terminology poses challenges to accurate coding. However, even for turns with no scientific terms, there is a gap between Human-Human and Human-GPT-4o accuracy. This suggests that while scientific terminology may contribute to reduced performance, the presence of scientific terminology does not fully explain the performance gap between GPT-4o and human raters on the science tasks, a gap which was not seen on the other 3 tasks.

**Table 5**

*Cohen's Kappa for Coding Performance on Chat Turns with and without Scientific Terms*

| Task | Contains Scientific Terms | # of chat turns | % of chat turns | Human vs. Human Kappa | Human vs. GPT-4o Kappa |
|---|---|---|---|---|---|
| Condensation | 0-no | 1231 | 81.58% | 0.784 | 0.573 |
| | 1-yes | 278 | 18.42% | 0.707 | 0.492 |
| | All | 1509 | 100% | 0.779 | 0.576 |
| Volcano | 0-no | 1383 | 91.59% | 0.697 | 0.629 |

| | | | | |
|---|---|---|---|---|
| 1-yes | 127 | 8.41% | 0.486 | 0.179 |
| All | 1510 | 100% | 0.685 | 0.604 |

The above finding naturally leads to Research Question 3 about the impact of coding frameworks on ChatGPT coding. Across the five tasks, the two science-related tasks were coded using Coding Framework 1, while the remaining three tasks used Coding Framework 2. We observed that ChatGPT models' coding performance on the two tasks using Coding Framework 1 was notably worse than that of human raters. In contrast, for the other three tasks using Coding Framework 2, ChatGPT models' performance was comparable to, or even better than, that from human raters. Since we have already ruled out the influence of scientific terms as the major factor, the most plausible explanation for this discrepancy lies in the coding frameworks themselves. This aligns with prior findings that the complexity and design of a coding framework can significantly influence ChatGPT's performance (Kocoń et al., 2023). Our results provide additional empirical support for this claim. Notably, Coding Framework 1 was developed primarily from a theoretical standpoint, whereas Coding Framework 2 was informed by both theoretical considerations and empirical data. ChatGPT appeared to have more difficulty interpreting and applying Framework 1, resulting in lower coding accuracy relative to human raters.

For Research Question 4 we focused on the two science-related tasks, as ChatGPT's performance on the other tasks was already comparable to or better than that of human raters. For the *Condensation* task, we first identified a subset of chat turns (1,265 out of 1,509, approximately 84%) where both human raters fully agreed, and treated this as the ground truth. We then analyzed the distribution of misclassified categories produced by GPT-4o (see left panel of Figure 4). Based on this analysis, we revised the prompt by explicitly calling out these miscoding categories, incorporating several miscoded examples and then instructed GPT-4o to re-code the same chat data. The updated results are shown in the right panel of Figure 4. Surprisingly, the overall Cohen's Kappa did not improve. Although the specific miscoded categories included in the revised prompt showed slight reductions, this improvement was offset by increased misclassifications in other categories. These results suggest that, for this task, the original prompt was already near-optimal, and that simply adding targeted feedback from miscoded cases does not yield meaningful gains in overall coding performance.

**Figure 4**

*Confusion Matrices for the Condensation Task based on Human Raters Agreed Chat Turns.*

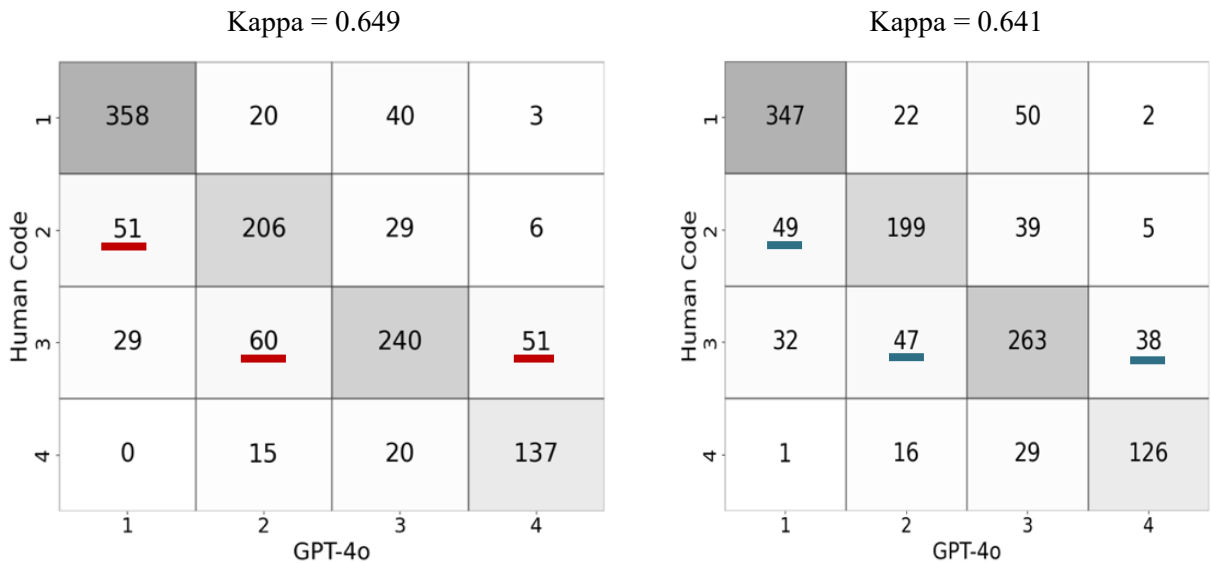

*Note.* These matrices are based on chat turns where human raters fully agreed (used as ground truth). **Left**: Results using the original prompt. **Right**: Results using the revised prompt, which incorporates examples from previously identified miscoded categories. Underlined numbers indicate the most frequently miscoded categories that were explicitly included in the updated

prompt. The labels 1 through 4 on the x- and y-axes correspond to the coding categories defined in Coding Framework 1 (see Table 1).

We conducted a similar analysis for the *Volcano* task, with the corresponding results presented in Figure 5. In this case, we observed an improvement in coding performance following the incorporation of previously miscoded examples into the prompt.

**Figure 5**

*Confusion Matrices for the* Volcano *Task based on Human Raters Agreed Chat Turns.*

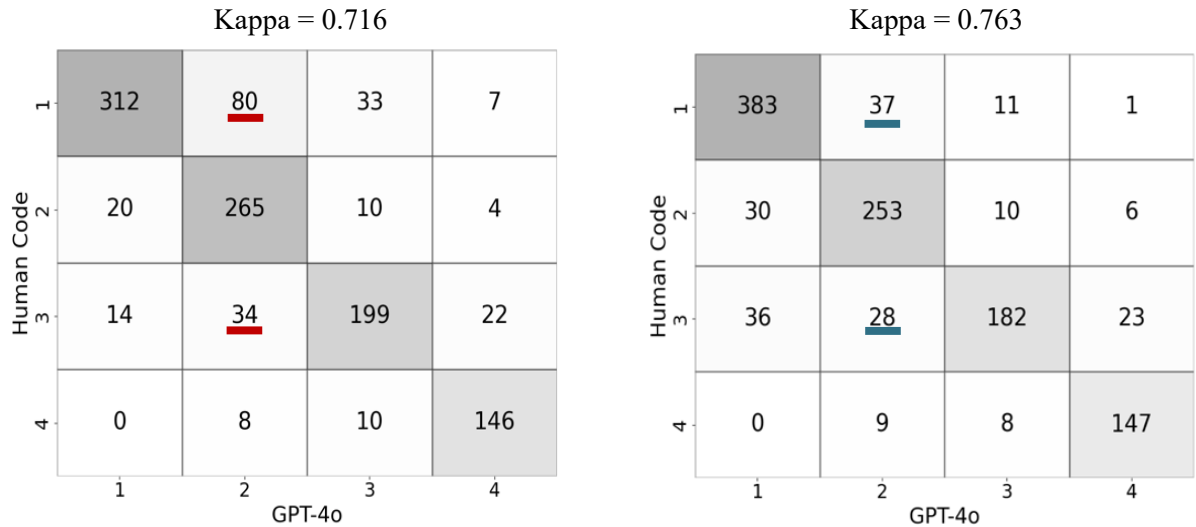

*Note*. These matrices are based on chat turns where human raters fully agreed (used as ground truth). **Left**: Results using the original prompt. **Right**: Results using the updated prompt, which incorporates examples from previously misclassified categories. Underlined numbers highlight the most frequently miscoded categories that were explicitly included in the revised prompt. The labels 1 through 4 on the x- and y-axes correspond to the coding categories defined in Coding Framework 1 (see Table 1).

Given the improvement observed on the subset of chat turns with full human agreement, we applied the updated prompt to recode the full *Volcano* dataset. The results are presented in Figure 6. A similar improvement in overall coding performance was observed, with Cohen's Kappa increasing from 0.604 to 0.637. However, this remains noticeably lower than the agreement level achieved between human raters.

**Figure 6.**

*Confusion Matrices for the Volcano Task Based on All Chat Turns.*

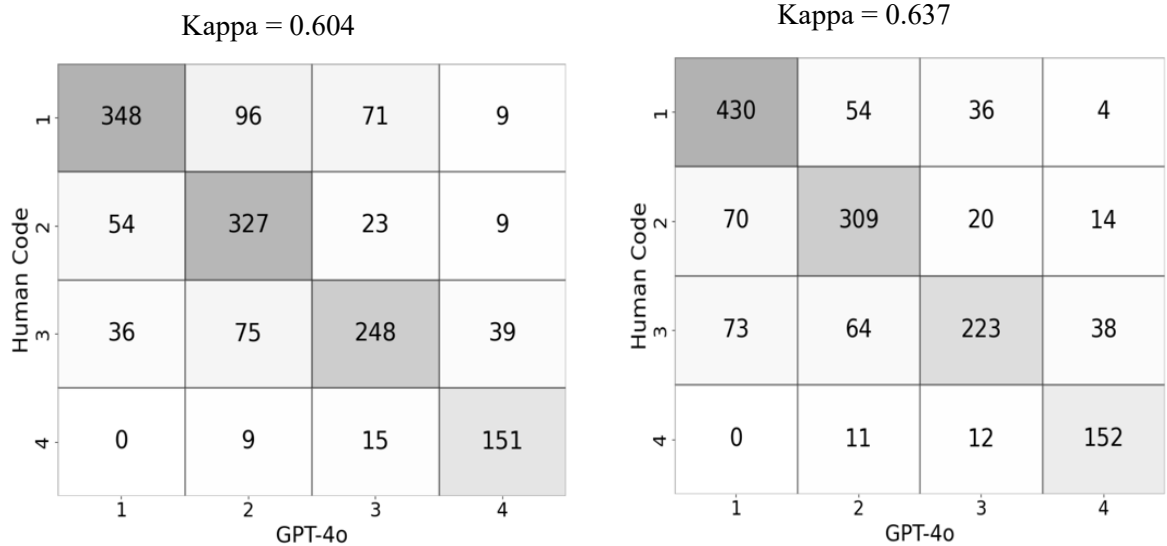

*Note.* These matrices are based on chat turns where human raters agree and disagree. **Left**: Results using the original prompt. **Right**: Results using the updated prompt, which incorporates examples from previously misclassified cases. The labels 1 through 4 on the x- and y-axes correspond to the coding categories defined in Coding Framework 1 (see Table 1).

**Discussion**

This study provides empirical evidence that ChatGPT can be effectively instructed to code communication data from CPS tasks under some typical coding frameworks. When the coding framework is relatively straightforward and data-driven, the performance of ChatGPT models can approach, match, or even exceed that of human raters (*exceed* occurs when the model matches the mean of human raters more closely than individual raters do). These findings support the feasibility of using ChatGPT to code communication data in CPS, offering a cost-effective and scalable way for assessing CPS.

Our findings also shed light on key questions underlying the use of LLMs for coding, as articulated in the four research questions that guided this study. First, we observed that all ChatGPT models achieved broadly comparable levels of coding performance, with GPT-4o as the most effective overall. Interestingly, the newer "reasoning" or "thinking" models, such as GPT-o1-mini and GPT-o3-mini, did not demonstrate improved coding accuracy, despite their strong performance on LLM benchmark leaderboards and higher computational costs. This result offers practical guidance for researchers and practitioners: when coding communication data from CPS tasks, selecting the most expensive or newest model does not necessarily yield better outcomes. Instead, models like GPT-4o may provide the best balance between performance and cost-effectiveness.

Second, we found that the nature of the communication content significantly affects ChatGPT's coding performance. Tasks involving complex and strategic interactions, such as negotiation, tended to result in lower agreement between ChatGPT and human raters. For instance, among the three tasks coded using Coding Framework 2, the *Negotiation* task consistently produced the lowest agreement for both human-human and human-ChatGPT comparisons. This task featured less structured, more open-ended communication, which likely made it more challenging for the model to interpret and classify interactions accurately. These findings underscore that both the content and the style of interaction play a critical role in determining the effectiveness of ChatGPT-based coding.

Third, we observed a clear contrast in ChatGPT coding performance between the two coding frameworks used in this study. Coding Framework 1, which was developed primarily from a theoretical perspective, proved more challenging for ChatGPT to apply accurately. In comparison, Coding Framework 2, which was informed by both theoretical foundations and empirical observations, resulted in significantly higher coding accuracy by ChatGPT. These findings align with prior research showing that the complexity and design of a coding framework can strongly influence how effectively LLMs interpret and apply it (Kocoń et al., 2023).

Finally, we observed mixed results when attempting to improve coding accuracy by incorporating feedback on common misclassifications into the prompt. For the *Condensation* task, we revised the prompt to include examples of frequently miscoded categories; however, this modification did not lead to an overall improvement in performance. While errors in the targeted categories were reduced, new misclassifications emerged in other categories, suggesting that the original prompt may have already been near optimal. In contrast, applying the same refinement strategy to the *Volcano* task led to a modest but consistent improvement in coding performance, with Cohen's Kappa increasing from 0.604 to 0.637. These findings suggest that feedback-driven prompt engineering can enhance LLM performance in some contexts, though its effectiveness may vary depending on task characteristics.

In applying ChatGPT to real-world coding tasks, we encountered several important practical considerations. One key issue involves the context window limitations of LLMs. All ChatGPT models used in this study have context window limits below 200,000 tokens, which made it infeasible to process the entire chat dataset in a single API call. To accommodate these constraints, we divided the data into batches of approximately 70 chat turns (typically corresponding to one complete collaborative session) per API request. Interestingly, this batching strategy not only addressed technical limitations but also enhanced reliability. When the number of turns exceeded 200, we observed increased response latency and a higher likelihood of mismatches between the number of chat turns submitted and the number of coded responses returned.

Additionally, we found that even with the model temperature set to zero and a fixed random seed, the model's outputs were not fully deterministic. Repeated API calls on the same dataset occasionally resulted in minor variations in coding decisions, though overall agreement statistics remained consistent across runs. We also found that averaging coding results from multiple runs (e.g., bootstrapping) did not lead to meaningful performance improvements.

Despite the promising results demonstrated in this study, several limitations should be acknowledged. First, the coding performance achieved with the prompts we have does not necessarily represent the upper bound; improved results may be attainable through further prompt refinement, fine-tuning, or alternative prompting strategies. Second, our findings are based on specific snapshot versions of the ChatGPT models. As LLMs continue to evolve, newer models, including those outside the GPT family, such as the latest versions of Gemini and Claude, may offer improved performance on such coding tasks. Third, due to logistical constraints, we were not able to apply both coding frameworks to all tasks. While this limitation does not affect our main conclusions, applying both frameworks across all tasks would offer a more complete basis for comparison. Fourth, the coding frameworks used in this study are still relatively simple and well-structured. Applying LLMs to more complex or nuanced frameworks may introduce additional challenges and will require further benchmarking to assess feasibility and reliability. Fifth, a critical concern in AI-based scoring is ensuring that outcomes are not biased against specific demographic groups (Jiang et al., 2024; Johnson & Zhang, 2024). Fairness in ChatGPT-based coding of communication data presents additional complexity compared to traditional automated scoring, as communication styles may inherently vary across demographic groups. We are curating new datasets to explore this issue and will report our findings in a future publication. Finally, we emphasize that although ChatGPT-like AI can code communication data based on its "understanding" of coding frameworks, it does not yet fully meet the validity standards required of human raters (American Educational Research Association, American Psychological Association, & National Council on Measurement in Education, 2014; Casabianca et al., 2025). Therefore, it should be used as a complement to human coding rather than a full replacement, at least until a new consensus on its validity is established.

In summary, our study provides robust empirical evidence supporting the use of ChatGPT for coding communication data to assist the assessment of CPS. While coding performance varies depending on the nature of the communication and the coding framework, our findings show that, under the right conditions, LLM-based coding can serve as a practical and scalable complement to human coding. These results also offer guidance for researchers and practitioners on selecting LLMs, engineering prompt, and designing coding frameworks.

**Acknowledgment**

This work was funded by the U.S. Army Research Institute for the Behavioral and Social Sciences (#W911NF-19-1-0106), the Education Innovation Research (EIR) program of the Department of Education (#S411C230179), and the ETS Research Institute. The views, opinions, and/or findings contained in this paper are those of the authors and shall not be construed as an official position, policy, or decision of the funding agencies, unless so designated by other documents.

**Appendix A: Example Prompts**

**Coding Framework 1**

Students form a team to work on a computer-based collaborative task. Students use the chat function on the computer to communicate with each other. We will code students' chats into four different categories. The four categories are:

Category 1: Chats that share information, resources and ideas. These chats bring divergent ideas into a collaborative conversation. Such as sharing individual responses to assessment items and/or pointing out relevant resources that might help resolve a problem.
Below is a list of example chats in this category:
["I feel Sam is correct",
"I thought 3",
"There is no change in mass.",
"what do you think, I am 100$ sure it is water molecules",
"I chose 3 and not certain about the second question",
"It's a closed system so the mass stays the same.",
'so i put that they were both moving',
"It's either C or A",
'Well condensation is when two temperatures are different, so it would be the cold can, cause the humidifier would make it warmer',
'water molecules move fast enough to escape ',
'I said the scale, because the scale takes the actual weight into consideration',
'I said because volume takes size into consideration']

Category 2: Chats that assimilate and accommodate knowledge/perspective taking. These chats help team's collaborative knowledge building and construction through negotiating with each other. Such as agreement/disagreement with each other, requesting clarification, elaborating/rephrasing other's ideas, identifying gaps, revising one's own idea.
Below is a list of example chats in this category:
["atoms cannot be destroyed",
"thermal contraction of solid mass",
"yes I said on the colder surface",
"I did put that the oil expands when heated..yay, maybe I know something",
"it is just steam; ie water",
'depends on how cold the window is',
"I'm thinking water molecules",
'We agree.',
"atoms cannot be destroyed",
"thermal contraction of solid mass"]

Category 3: Chats that regulate problem-solving activities. These chats focus on the collaborative regulation aspect of the team discourse, such as identifying goals, evaluating teamwork, checking understanding.
Below is a list of example chats in this category:
["I don't know if all molecules move at the same speed if the liquid is the same temperature",
"I think they move at different speeds but not 100%",
"Do they not break down?",
"so the pan gets smaller?",
"I can't tell, but it looks like the molecules are moving at the same speed for a warmer can",
"are you good at science?",
"I dislike that you have to manually scroll the chat box.",
"I didn't think iron would shrink...",
"I thought it expanded because it's heated...guess I know nothing about science.",
"I like it"]

Category 4: Chats that maintain a positive communication atmosphere. These chats are content irrelevant social communications. They try to keep the collaboration going, but not related to the task content. They help to keep social communications, such as greeting and pre-task chit-chat, emphatic expression of an emotion or feeling.
Below is a list of example chats in this category:
['hi',
"Hi, I'm Jake",
"where are you from?",
'Yea me neither to be honest',
'Wow',
"Wait you're a real person?",

"Are you a bot?",
'ummmmm just wing it',
':)',
"BOOM",
"wow i'm so tired hope this is the last set of questions",
'I stand by the picture haha',
"Indeed"]

Below are the turns of chats. Please assign a category number to each of them and return the coding list. Do not include the original chats.

1. I think they are both moving but the solid one moves slower
2. I thought that atoms didnt move in a solid, but I'm wasnt totally sure
3. I'm not sure either
4. We can go with your answer
5. ... ...

**Coding Framework 2**

Students form a team to work on a computer-based collaborative task. Students use the chat function on the computer to communicate with each other. We will code students' chats into five different categories. The five categories are:

Category 1:     Chats that help maintain communication. Such as greeting and pre-task chit-chat, emphatic expression of an emotion or feeling, chat turns that deal with technical issues, and chat turns that don't fit into other coding categories (generally off-topic or typos).
Below is a list of example chats in this category:
["Hi, I'm Jake",
"where are you from?",
"hi, can you see this?",
"lol",
";-)",
"so",
"I'm terrible at this.",
"BOOM",
"Are you a bot?",
"my screen froze",
"my submit button is grayed out",
"who doesn't like discounts, right?"]

Category 2: Chats that focus discussion to foster progress. Such as Making Things Move or Monitoring time to stay on track (not related to task content), steering conversation back to the task.
Below is a list of example chats in this category:
["ah there's 8 minutes",
"let's get this done.",
"hit next",
"we should focus on the task",
"let's talk about the next one",
"what are we supposed to do?",
"We're waiting on Sue to submit",
"I am ok lets move on"]

Category 3: Chats that ask for input on the task. Such as asking for information related to the task, asking for task strategies (regardless of phrasing, no strategy proposed), and asking for task-related goals or opinions. These chats are usually questions and end with a question mark:'?'.
Below is a list of example chats in this category:
["What did you put for best?",
"Do you know what any of the letters are?",
"what about where we'll present?",
"How do I do this?",
"What did you try?",
"Should I add J+J?",
"Why did you put A for best?",
"Why don't you like the raffle?",
"Why are you only adding two numbers?",
"So how about ABC if we all agree?",
"C B A?",
"cba?",
"Can we ...... ?",
"How about ......"]

Category 4: Chats that contribute details for working through the task. Such as sharing information related to the task, sharing strategies for solving the task, and sharing task-related goals or opinions.
Below is a list of example chats in this category:
["Utilities are included in A",
"C=3",
"It's alright, we still know A is 5",
"Let's try adding three numbers",
"If A is 1 then B must be 2",
"If utilities are included, it means the rent is higher",

"I had BCA",
"I want the raffle",
"I think we should do it outside.",
"I don't like paying a pet deposit",
"I'm submitting ACB, since we all agreed.",
"ah, i hate that"]

Category 5: Chats that Acknowledge partner(s) input and may continue off that input with their own. Such as neutrally acknowledge a partner's statement, agree with or support a partner's statement, disagree with a partner's statement, adding details to a previously made chat turn (their own or another player), suggesting a solution of give and take (e.g., if you give me this, I'll give you that. May simultaneously reject an option on the table and propose a compromise). Below is a list of example chats in this category:
["okay, thanks",
"we nailed it",
"we got all!",
"Good",
"You make a fair point",
"i'll try it",
"Oh I see. That's a good reason.",
"Yeah I wish B had an elevator",
"Alright, same for me",
"weekends are good",
"I don't think so.",
"can't do afternoons, how about evenings?",
"mine doesn't say that, but I like that A has two bathrooms",
"Good job!"]

Below are the turns of chats. Please assign a category number to each of them and return the coding list. Do not include the original chats.

1. Hello all
2. Hi all!
3. First task, location. we all must agree on a location to "win"
4. I would choose inside due to weather concerns, or virtually. Anyone else have any thoughts?

**Appendix B: Scientific Terms**

a/c, absolute zero, accuracy, accurate, acoustics, activity, ai, air, air conditioning, air molecules, alert, alert level, alert levels, algebra, analogy, aquarium, area, artificial intelligence, atomic, atomic mass, atoms, audio, balance scale, balloon, beneficiary, boil, boiled, boiling, boiling point, bubbles, cancer, cell phone reception, cell phone service, cell reception, central a/c, central air, central air conditioning, chain of events, chamber, chamber pressure, chart, chemistry, closed system, code violations, cold, cold can, computer simulations, condensation, condense, condensed, condenses, condensing, contracts, cool, covid, crater, d = m / v, data, date range, defensible, degrees, dense, densely packed, density, developmental support, distilling, early intervention, earth, earthquakes, electric cars, electrons, element, elimination, energy, equation, equations, erupt, eruption, eruptions, evaporate, evaporated, evaporates, evaporation, expand, expands, expansion, experimenting, fireplace, fog, freeze, frequencies, frequency, fresh water, froze, gas, gas state, gaseous, gravity, ground movement, handicap accessible, health, health code violations, heart, heat, heat energy, heated, heating, hermetically sealed, hf, hf activity, hf events, hf readings, high, high frequency, high frequency events, high frequency waves, hot air, humid, humidity, hydrogen, hypothesis, ice, impurities, internet connectivity, iron atoms, junior scientist, kitchen cleanliness, lab, lake, lava, lava flow, lf, lf events, liability, liability insurance, lighting, liquid, low, low frequency, low frequency events, magma, mass, material, matter, med, medical, medium, metal, microwave, mobility, moisture, molecule, molecules, movement, natural light, negative, neutrons, objective, ocean, ocean water, oil, oxygen, pandemic, parasite, particles, physical, physics, point value, predict, pressure, process, pure water, relative humidity, rocks, rocks cracking, salinity, salt, salt particle, salt particles, salt-water aquarium, saltwater aquarium, sample, scale, science, scientific, scientifically, security deposit, security door, seismic, seismic activity, seismic events, seismometer, seismometers, sequence of events, series of events, simulation, simulator, slowed down, snowstorm, solid state, solute difference, sound absorption, speech, speed, speeds, stability, stable, stable period, stable tremors, state, steam, superimposing, surface, temp, temperature, temperature, temperature control, temperature difference, temperatures, theory, thermal expansion, tremor, tremors, unknown, vapor, variables, video, viscous, volcanic, volcanic seismic events, volcano, volume, warm, warm air, washer and dryer, water, water droplets, water drops, water molecules, water particles, water vapor, weighing, weight, wi-fi, wooden

Jiang, Y., Hao, J., Fauss, M., & Li, C. (2024, July). Towards fair detection of AI-generated essays in large-scale writing assessments. In *International Conference on Artificial Intelligence in Education* (pp. 317-324). Cham: Springer Nature Switzerland.

Johnson, M., & Zhang, M. (2024). Examining the responsible use of zero-shot AI approaches to scoring essays. *Scientific Reports*, *14*(1), 1-10.

Jurafsky, D & Martin, J. H. (2019). Speech and Language Processing (3rd ed. draft). https://web.stanford.edu/~jurafsky/slp3/

Kocoń, J., Cichecki, I., Kaszyca, O., Kochanek, M., Szydło, D., Baran, J., ... & Kazienko, P. (2023). ChatGPT: Jack of all trades, master of none. *Information Fusion*, 101861.

Koschmann, T. D. (1996). *CSCL: Theory and practice of an emerging paradigm*. New York, NY, USA: Routledge.

Kyllonen, P., Hao, J., Weeks, J., Kerzabi, E., Wang, Y., and Lawless, R., (2023). Assessing individual contribution to teamwork: design and findings. Presentation given at NCME 2023, Chicago, IL., USA.

LMArena. (n.d.). *Chatbot Arena Leaderboard*. LMArena. Retrieved April 8, 2025, from
https://lmarena.ai/?leaderboard

LLM Stats. (n.d.). *LLM Leaderboard 2025: Verified AI Rankings*. Retrieved April 8, 2025, from

Liu, L., Hao, J., von Davier, A. A., Kyllonen, P., & Zapata-Rivera, J. D. (2016). A tough nut to crack: Measuring collaborative problem solving. In *Handbook of research on technology tools for real-world skill development* (pp. 344-359). IGI Global.

Liu, L., Rogat, A., & Bertling, M. (2013). A CBAL™ science model of cognition: Developing a competency model and learning progressions to support assessment development. *ETS Research Report Series*, *2013*(2), i-54.
42

43